# Motion of small bubbles and drops in viscoelastic fluids

D. Langevin

Laboratoire de Physique des Solides, UMR CNRS 8502, Université Paris Sud

## Abstract

Viscoelastic materials containing bubbles or drops are encountered in numerous application fields, and are presently the object of much interest. The motion of bubbles and drops in these matrices can be significantly different than in Newtonian fluids. This review is restricted to the case of motion in quiescent fluids (or small Reynolds number) and of small bubbles and drops, that are not appreciably deformed during their motion. It includes a brief description of properties of viscoelastic systems and of the motion of solid particles in these systems. The case of very small drops undergoing Brownian motion is related to recent advances in microrheology. The motion of larger drops and bubbles due to gravity in yield stress fluids is discussed and linked to Ostwald ripening. Recent advances on the understanding of the rheological properties of the composite systems are also briefly discussed.

## 1. Introduction

Non-Newtonian fluids containing bubbles or drops are encountered in numerous application fields, such as foods, paints, cosmetics, medicine, bioreactors, oil and gas exploration and construction materials [1]. For instance, microbubbles for medical imaging are dispersed in biological fluids such as blood, which are almost always non-Newtonian. Bubbles can be incorporated into non-Newtonian concentrated dispersions of drops to produce foamed emulsions, with also many applications, for instance in the road industry [2].

In quiescent fluids, small bubbles and drops are submitted to Brownian motion and to buoyancy forces, due to the difference between their density $\rho_i$ and that of the fluid $\rho$. When the drops are submicronic, Brownian motion dominates, whereas larger drops either sediment if $\rho_i > \rho$ or rise if $\rho_i < \rho$. Bubble sizes are usually above tens of microns, and bubbles rise in the fluid. When the bubbles or drops are large, their velocities V are large, they deform and do not remain spherical: this happens when the capillary number Ca = $\gamma V/\eta$ is large, $\gamma$ being the surface tension and $\eta$ the viscosity of the continuous phase. The paper will be limited to the cases of small bubbles/drops that remain spherical, and of drops made of a Newtonian fluid, only the dispersing fluid being non-Newtonian.

The Reynolds numbers involved in this type of motion are small, especially with non-Newtonian fluids, generally quite viscous. The topic has been extensively investigated with Newtonian fluids [3]. The motion depends in principle on the boundary conditions at the bubble/drop surface. Although bubbles/drops are fluid, it is currently observed that they behave as solid particles. This is due to residual contamination or to surface active species, when present, that move along the surface during the motion, creating surface tension gradients that immobilize the surface [4]. Because of this condition, much has been learned from the behavior of small solid particles, that were extensively studied. In particular, the knowledge of their behavior in Newtonian fluids is quite advanced.

The motion of particles in non-Newtonian fluids remain much less well known. In this paper, several recent advances will be described. The behavior of non-Newtonian fluids will be presented first, followed by a brief description of the knowledge achieved with solid particles and a discussion on the peculiarities of bubbles and drops motion.

## 2. Non-Newtonian fluids

Non-Newtonian fluids, also called viscoelastic fluids, are fluids which viscosity $\eta$ is not constant. In the linear regime, $\eta$ depends for instance on the (angular) frequency $\omega$ of a sinusoidal shear applied in a rheometer. This occurs when the change in local interactions in the fluids produced by the shear relaxes in a timescale comparable to $1/\omega$. Because of the causality principle and in order to comply with the Kramers-Kronig relations, these fluids possess a finite shear modulus G' and behave as solids above a certain frequency. The modulus G' is called "storage modulus", whereas the product frequency times viscosity G" is called "loss modulus" [5]. Frequency variations of the storage and loss modulus for a fluid with a well-defined relaxation time $\tau$ are shown in Figure 1.

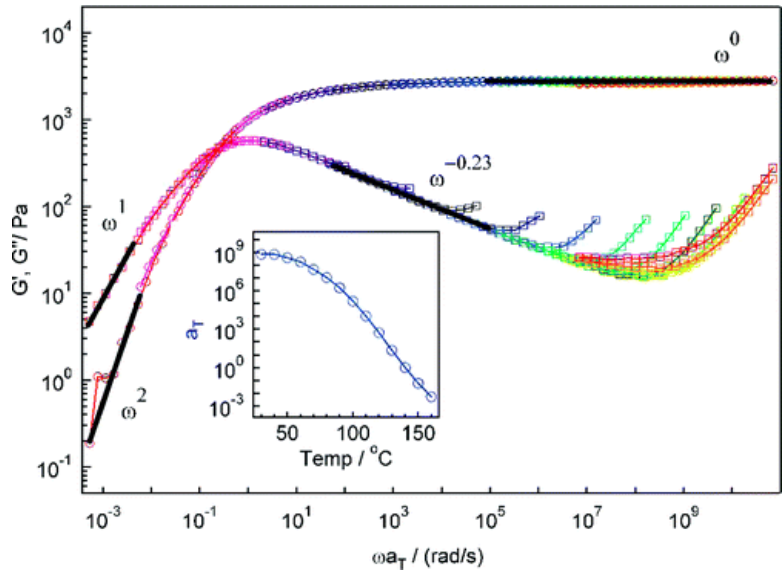

Figure 1. Storage (G', circles) and loss (G'', squares) modulus versus frequency for a an hydrogen-bonded supramolecular polymer networks of associating block copolymers. The colors for each plot correspond to different temperatures. The temperature-dependent shift factors, $a_T$, are plotted in the inset. Reprinted with permission from [6]. Copyright 2008 American Chemical Society.

The frequency variations are described by the Maxwell model:

$$G'(\omega) = G_0 \frac{(\omega\tau)^2}{1+(\omega\tau)^2} \quad and \quad G''(\omega) = G_0 \frac{\omega\tau}{1+(\omega\tau)^2} \qquad (1)$$

where $G_0$ is the high frequency modulus (for $\omega \gg \tau^{-1}$). G' and G'' cross at a frequency equal to $\tau^{-1}$. The deviations to the Maxwell model (G'' increasing with frequency) seen at larger frequencies in Figure 1 are due to faster relaxations.

Non-Newtonian fluids also exhibit non-linear rheological behavior. In a simple shear flow with a velocity $V_x$ along x varying along y: $V_x = y\,\dot{\gamma}$, $\dot{\gamma}$ being the shear rate. The viscous stress tensor is such as $\sigma_{xy} = \eta\,\dot{\gamma}$. The viscosity $\eta$ of these fluids can decrease or increase when $\dot{\gamma}$ is increased; the fluid is then said to be "shear-thinning" or "shear thickening", respectively (see Figure 2).

Another viscosity can be defined in an extensional flow with a velocity $V_x$ varying along Ox: $V_x = x\,\dot{\varepsilon}$, $\dot{\varepsilon}$ being the extensional rate. The extensional viscosity is denoted $\eta_E$, such that $\sigma_{xx} = \eta_E\,\dot{\varepsilon}$. For Newtonian fluids $\eta_E = 3\eta$. In solutions of long polymers, the extensional viscosity can be very large due to the dissipation associated to the extension of the polymer chains.

When the storage modulus is larger than the loss modulus, in principle the medium cannot flow. This is the case for a number of the viscoelastic fluids that will be considered in this paper. However, above a critical stress $\sigma_Y$ called yield stress, or a critical strain call yield strain, the solid melts, after which G'' remains larger than G'. The yield strain is frequently about 10 %.

A commonly used empirical expression of the shear stress $\sigma$ ($\sigma_{xy}$ here) was proposed by Herschel and Buckley:

$$\sigma = \sigma_y + \alpha\,\dot{\gamma}^n \qquad (2)$$

where $\alpha$ is a constant and n an exponent: n < 1 for shear thinning fluids and n > 1 for shear thickening fluids. For Newtonian fluids, $\sigma_y$ = 0, n=1 and $\alpha$ = $\eta$. Figure 2 illustrates different typical behaviors.

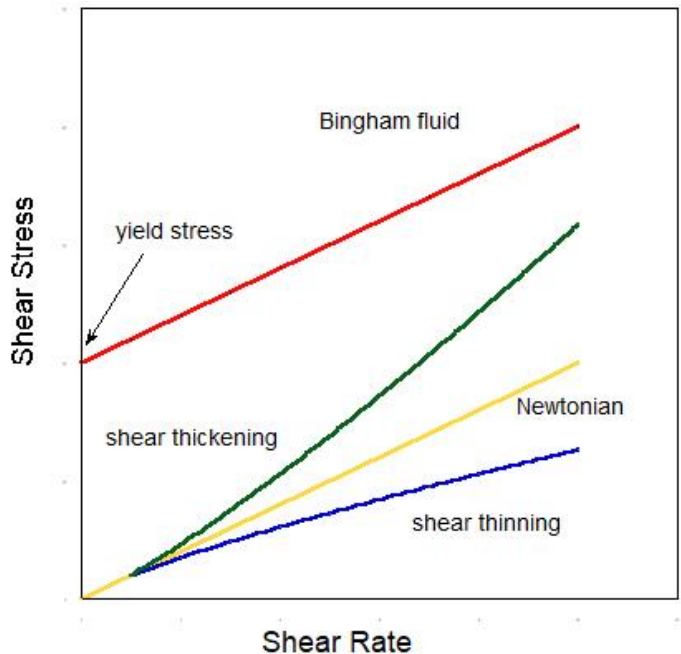

Figure 2. Scheme of typical non-Newtonian behavior. Red curve: Bingham fluid, $\sigma_y \neq 0$, n = 1. Fluids with no yield stress: yellow curve, Newtonian fluid, n = 1; blue curve, shear thinning (pseudoplastic) fluid, n = 0.8; green curve, shear thickening (dilatant) fluid, n = 1.2

Normal stress coefficients can be defined by dividing the differences between normal stresses $\sigma_{xx}-\sigma_{yy}$ and $\sigma_{xx} - \sigma_{zz}$ by $\dot{\gamma}^2$.

## 2.1. Polymer solutions and melts

The best understood behavior of non-Newtonian fluids is that of polymer solutions and melts [7][8]. As stated by Cates and Fielding, microscopic models of polymer rheology arguably represent one of the major intellectual triumphs of 20th century Statistical Physics [9]. Polymers are very long molecules, chain-like, that adopt coil configurations in dilute solutions. They remain straight along a distance b called "Kuhn length" and start bending at larger distances due to Brownian motion. Note that coiling is somewhat described using the notion of "persistence length", equal to b/2.

If the number of monomers N of the polymer chain is larger than Ne (of order 20, depending on the polymer), the chains can entangle: a given chain is surrounded by other chains that form a tube within which the given chain will move. The diameter of the tube in a polymer melt is a ≈ b $Ne^{1/2}$ and is larger than b.

The smallest relaxation time is the so-called "reptation" time $\tau_{rep}$, needed for the chain to exit the tube. Shorter relaxation times are also present, and associated to smaller scale chain motion. At long times, or small frequencies, the relaxation is characterized by the reptation process and the rheological behavior is rather well represented by the Maxwell model (Equation 1) with $\tau = \tau_{rep}$.

Figure 1 showed an example of the frequency variation of the storage and loss moduli. The data taken at different temperatures has been rescaled, confirming the existence of a single relation time. This time is equal to the inverse frequency at which G' = G'' ($\tau_{rep}$ = 1/$a_T$ in Figure 1).

At low frequencies, the Maxwell model usually describes qualitatively well the data. At high frequencies, the medium is solid-like (G' > G'') and the loss modulus deviates from Equation 1 due to the existence of relaxations at shorter timescales. The storage modulus is rather constant with frequency, in the range of the so-called «rubbery plateau».

P.G.de Gennes used scaling laws to predict that $\tau_{rep}$ should scale as $N^2$, N being the number of monomers of the chain, and that the viscosity should scale as $N^3$, not far from the experimental values $N^{3.4}$. The differences are thought to originate from fluctuations in tube length.

The behavior of entangled polymer solutions is similar. In dilute solutions, the polymer chains take a coil configuration. When the polymer concentration C is increased, the coils become in contact at a concentration C = C*. The microstructure resembles a mesh, with a mesh size $\xi$ equal to the coil radius of gyration at C* and decreasing with C as $\xi \approx b\, C^{-0.76}$ in good solvent conditions, and as $\xi \approx b\, C^{-1}$ in poor solvent conditions or in concentrated solutions (C being expressed in terms of volume fraction). The mesh size $\xi$ is a correlation length beyond which interactions between chains are screened.

At a slightly higher concentration $C_e$, the tube diameter becomes equal to the coil size, the chains begin to entangle. The diameter of the tube is a ≈ b $Ne^{1/2}$ $C^{-0.76}$ in good solvents, and a ≈ b $C^{-2/3}$ in poor solvents or in concentrated solutions. The tube diameter is therefore always larger than the Kühn length. In the entangled regime, the viscosity increases with concentration with power law exponents of 3.9 and 4.7 for good and poor solvents respectively, in reasonable agreements with experiments (somewhat larger exponents are found in experiments, as with polymer melts and probably for the same reasons).

Polyelectrolytes are polymers bearing electrical charges. Their solution behavior is very different from that of neutral polymers. Their persistence length contains an intrinsic contribution from the chain flexibility, similar to that of neutral polymers and an electrostatic contribution of the order of the Debye screening length. At low ionic strength, the electrostatic contribution dominates and the coil radius is very large. The concentration C* is therefore very small, but due to the large persistence length, the chain cannot entangle, they simply overlap. This leads to a viscosity that varies as $C^{1/2}$, much more slowly with concentration than for neutral polymers. As C increases, the ionic strength increases and the persistence length decreases, until the electrical charges are screened and the polymer behaves as a neutral polymer. This behavior is also observed if sufficient salt is added, as for instance in biological conditions when the solutions contain large amounts of salts (> 0.1M). Biopolymers such as DNA or actin (chain of proteins that assemble or disassemble under the action of ATP) also have a very large intrinsic persistence length.

Polymer-like systems scan be formed by surfactant aggregating into very long micelles, called « giant » or "worm-like" micelles. The micelles continuously break and reform, with a characteristic time $\tau_b$. The effective reptation time is $(\tau_{rep}\tau_b)^{1/2}$, where $\tau_{rep}$ is the reptation time corresponding to the average micelle length. The rheological behavior is arguably the closest to the Maxwell model[9].

## 2.2 Particle dispersions

Dilute dispersions of particles behave as Newtonian fluids [5]. At small volume fraction $\phi$, the viscosity increases with $\phi$ and is usually well represented by the Krieger formula:

$$\eta = \eta_c \left(1 - \frac{\phi}{\phi^*}\right)^{2.5\phi^*} \tag{3}$$

The limit fraction $\phi^*$ is the close packing volume fraction beyond which further particles cannot be added. For hard spheres dispersions, $\phi^* \sim 0.64$, which is the random packing volume fraction of spheres. Equation 3 also predicts that the viscosity is independent of particle size. When there are attractive forces between particles, $\phi^*$ is smaller and the viscosity increases faster with drop volume fraction. In some systems, the particles can start aggregating at very small $\phi$ (a few percent) and gels can be obtained for instance with silica particles (silica gels). Depending on the type of particles and on the interactions between them, the volume fraction at which gels form can be quite different. Gels can be also obtained with non-spherical particles such as clays. The variation of G' and G" with frequency is similar to that of polymers, with a characteristic time $\tau$ being a structural relaxation time.

## 2.3 Emulsions and foams

Emulsions and foams are collections of drops and bubbles [10]. Dilute dispersions of bubbles (bubbly liquids) are very unstable because the bubbles rapidly rise at the top of the dispersion. Dilute emulsions can be studied more easily. The viscosity of dilute emulsions is well described by the Krieger equation, but $\phi^*$ is much larger than 0.64 at which the drops come into contact. Above this volume fraction, the drops deform and the emulsion becomes viscoelastic. Note that this can happen earlier if the ionic strength is large and if the drops cluster, or if the drops are very small and bear electrical charges, in which case the volume fraction should also include the cloud of counterions.

The frequency variation of G' and G'' is similar to that shown in Figure 1 for polymers, both for emulsions and foams. The structural relaxation time appears to be the time between reorganization events of drops and bubbles assemblies during which neighbors are exchanged (called T1 events in foams).

## 2.4 Non-linear behavior

Polymers frequently strain-soften, possibly because the shear increases the tube length: this increase is rapidly relaxed by the motion of the free ends and the retraction suppresses a fraction of the tube segments. Retraction cannot occur in cross linked polymer networks, which indeed do not strain soften [9]. Solutions of worm-like micelles have a peculiar shear thinning behavior. At small shear rates, the shear stress is linear in $\dot{\gamma}$ and reaches a plateau at a certain shear rate, meaning that in a large range of shear rates, the viscosity is inversely proportional to the shear rate. This behavior is believed to be associated to "shear banding": the micelles partially align in layers with different strain rates but equal stress, the normal of the layers being parallel to the velocity gradient [9]. Shear banding is observed in many other viscoelastic systems and care has to be taken in using the viscosity values given by a rheometer in which uniform velocity profiles are assumed to process the data.

The nonlinear rheological properties of other non-Newtonian fluids are also very diverse. As polymer solutions, most colloidal suspensions show a simple yielding behavior, where both the storage and loss modulus decrease monotonically as the strain increases. Less common is strain hardening (or shear thickening), where the moduli increase. There is also an intermediate behavior where G' continuously decreases above a critical strain, whereas G'' exhibits a peak before decreasing at even larger applied strains. This behavior is ubiquitous in metastable complex fluids such as hard or soft sphere suspensions, emulsions and foams. Weitz and coworkers showed that this behavior can be explained by assuming that the structural relaxation time τ decreases when the shear rate increases as:

$$\frac{1}{\tau} = \frac{1}{\tau_0} + \alpha \dot{\gamma}^\nu \qquad (4)$$

$\alpha$ being a constant and $\nu$ a positive exponent close to 1 [11]. For a Maxwell viscoelastic fluid, Equation 1 can be used to show that a given frequency ω >> $1/\tau_0$, the variation of G' and G'' with shear rate mirrors its frequency variation. Weitz and coworkers proposed that investigating frequency variations at constant shear rate is a useful method to extend the range of frequencies probed, provided Equation 4 holds. They called the method "strain-rate frequency superposition" (SRFS). Note that the method is not applicable when the deviations to linear behavior are important and it should therefore be used with caution [12].

### 3. Motion of solid particles in viscoelastic fluids

#### 3.1 Brownian motion

The motion of small particles in complex fluids was envisaged long ago as a tool to probe fluid microstructure. For instance, it was conjectured that particles smaller than the mesh size ξ of a polymer semi-dilute solution should move as in the pure solvent, whereas particles larger than ξ should move as in a medium of viscosity equal to the macroscopic viscosity. P.G.de Gennes proposed that the viscosity 'seen' by the particle should scale as exp(d/ξ), d being the particle diameter [13]. Experiments were performed by looking at the sedimentation of spheres of various sizes in a centrifuge and indeed the prediction by de Gennes was confirmed [14]. In these experiments, the sedimentation velocity V is measured and the friction coefficient f exerted by the fluid on the particles is determined, the friction force F = f V being equilibrated by the centrifugal force; for Newtonian fluids and spherical particles of diameter d, f = 3πηd. Studies of the Brownian motion of small particles in polymer networks were also performed using light scattering methods. In these experiments, the diffusion coefficient D of the particles is measured, and is equal to $k_BT/f$, $k_B$ being the Boltzmann constant and T the absolute temperature.

It was shown later by Rubinstein and coworkers that the motion of particles in semi-dilute polymer solutions is rather complex, due the different characteristic times in the medium [15]. Figure 3 shows the time variation of the mean square displacement $<r^2(t)>$ multiplied by the particle radius. This product is linear in time for a Newtonian fluid, with a slope equal to 3Dd = $k_BT/(\pi\eta)$. Small particles (d < ξ) feel the solvent viscosity (dash-dotted line in Figure 3). When d is larger than ξ but smaller than the tube diameter, a sub-diffusive motion is expected $<r^2(t)> \sim t^{1/2}$. Particles of size slightly larger than the tube diameter move due through hopping process between different tubes and D ~ exp (-d/a). Because the overlap and entangling concentrations differ only by a factor $\sqrt{N_e}$, and scales with concentration like ξ in good solvents the behavior observed earlier in [14] can be rationalized. Larger particles are first immobilized until the reptation time is reached, after which they move, the friction coefficient becoming controlled by the macroscopic viscosity.

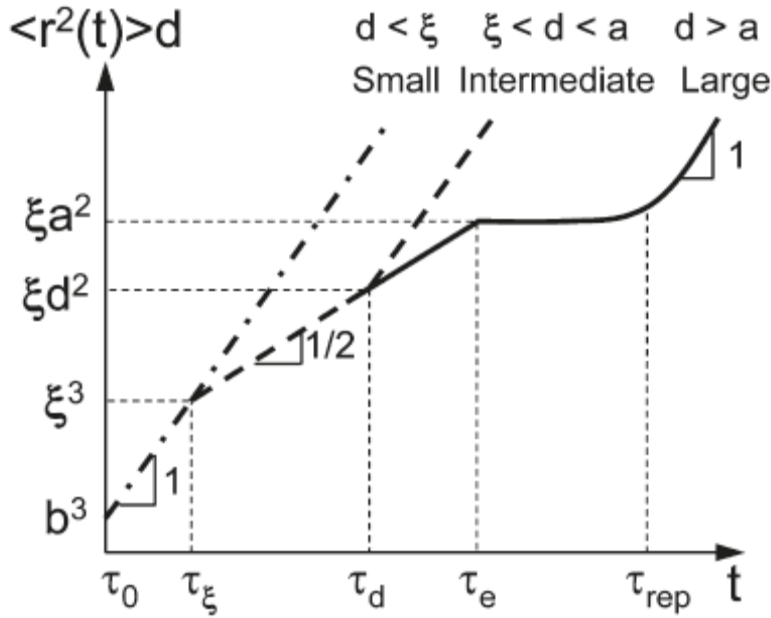

Figure 3. Time dependence of the product of mean-square displacement $\langle \Delta r^2(t) \rangle$ and particle size $d$ for small particles ($b < d < \xi$, dash-dotted line), intermediate size particles ($\xi < d < a$, dashed line), and large particles ($d > a$, solid line) in polymer solutions ($\xi \simeq b$ in polymer melts). Here $\tau_0$ is the relaxation time of a monomer, $\tau_\xi$ is the relaxation time of a correlation domain, $\tau_d$ is the relaxation time of a polymer segment with size comparable to particle size $d$, $\tau_e$ is the relaxation time of an entanglement strand, and $\tau_{rep}$ is the relaxation (reptation) time of a whole polymer chain. Logarithmic scales. Adapted with permission from [15]. Copyright 2011 American Chemical Society

In permanent networks, the motion is controlled by obstruction effects. For instance, an early experiment of fluorescence recovery after fringe pattern photobleaching (FRAPP), the motion of a fluorescent molecule in silica gels was studied. The diffusion coefficient decreased from that in the pure solvent $D_0$ at short fringe spacing (equivalent to short times) to smaller values at large fringe spacing (equivalent to long times): $D = D_0/T_{ort}$, $T_{ort}$ being the tortuosity of the medium. The crossover takes place when the fringe spacing is comparable to the mesh size of the silica gels [16].

The Brownian motion of particles larger than the characteristic size of the microstructure of the medium gives information about the local friction. This idea was used to develop particle tracking techniques and the method is now known as micro-rheology [17]. In "passive" measurements, the complex modulus is deduced the Brownian motion of particles using a generalized Stokes Einstein relation and a Laplace transform of $<r^2>$:

$$\tilde{G}(s) = \frac{2k_B T}{\pi d s <\tilde{r}^2(s)>} \qquad (5)$$

where s is the Laplace frequency. This equation is valid in systems at thermal equilibrium, but not in glasses, in active systems such as motor proteins or in externally driven systems. Is not valid either at large stresses or strains, when the response becomes non-linear. Equation 5 can be also used with fluid particles, in which case the friction coefficient depends on the limit conditions at the particle's surface: $3\pi\eta d$ for no-slip condition ad $2\pi\eta d$ for mobile surfaces (see section 4).

Diffusing wave spectroscopy also allows studying Brownian motion of particles. It provides access to high frequency linear viscoelasticity and is still widely used. When Equation 5 is not applicable, or when the viscoelastic moduli are too high, active micro-rheology can also be performed using magnetic or optical tweezers.

In general, micro-rheology measurements are consistent with those obtained with standard bulk rheometers, provided the particles used in the micro-rheology experiments are significantly larger than the characteristic sizes of the medium microstructure. Standard rheometers still lack the ability to probe spatial heterogeneities, which is why micro-rheology is widely used in biological samples such as cells [18]. Standard rheometers are also limited to low frequencies (~100 Hz as compared to $10^6$ Hz for micro-rheology), due to the inertia of their components. They may also disrupt solids with a small modulus G', such as for instance weak hydrogels.

Let us mention that imaging of Brownian motion of very small particles was recently demonstrated using an electron microscope equipped with a "liquid cell" enclosed in graphene sheets [19]. Imaging by this method in complex environments such as biological cells has been used to monitor the growth or the dissolution of the particles [20].

## 3.2 Gravity-driven motion

Larger spheres are sensitive to gravity which dominates over Brownian motion. At small Reynolds number, the velocity V is given by the Stokes expression:

$$V = V_{St} = \frac{\Delta\rho \, g \, d^2}{18 \, \eta} \qquad (6)$$

$\Delta\rho$ being the density difference between the particle and the surrounding fluid, $\eta$ its viscosity and g the gravity constant. A drag coefficient is generally used, which is such as:

$$C_D = \frac{8F}{\rho\pi d^2 V^2} = \frac{24}{Re} \qquad (7)$$

where Re is the Reynolds number defined as:

$$Re = \frac{\rho V d}{\eta} \qquad (8)$$

where $\rho$ is the fluid density. Experiments with solid spheres falling in Newtonian fluids have confirmed these results when Re is small.

Non-Newtonian fluids are usually viscous, and when their viscosity is of the order of say 1 Pa.s or larger, the longest relaxation time $\tau$ can be of the order of seconds. Strong non-linear effects can be observed at low shear rates, $\dot{\gamma}$ ~ 1 s$^{-1}$. This relaxation time $\tau$ is related to another a dimensional quantity, the Deborah number:

$$De = \frac{\tau V}{d} \qquad (9)$$

When the fluid is predominantly elastic (G' > G''), the particle does not move excepted if the force that it exert on the liquid per unit area is larger than the yield stress

The motion of single spheres in these fluids is quite different from that in Newtonian fluids, due to both shear-thinning and normal stresses. For instance, the velocity in the wake can invert direction and the fluid can move toward the particle. When more than one particle is present, attractive forces can arise. When many particles are present, they usually align in strings. These features are not yet fully understood [1].

## 4. Motion of bubbles and drops in viscoelastic fluids

Understanding the behavior of bubbles and drops in less easy than for solid particles. Indeed, bubbles and drops can deform if they are large enough. Bubbles remain spherical if both the inertial and viscous forces are small compared to surface tension forces [21]. This happens when the Weber number *We* and the capillary number *Ca* are small :

$$We = \frac{\rho V^2 d}{2\gamma} \qquad Ca = \frac{\eta V}{\gamma} \qquad (10)$$

Small bubbles and drops move very slowly, either due to Brownian motion if they are submicronic, or due to gravity if they are larger. For instance, with bubbles of diameter of 0.5 mm in water that rise due to buoyancy, *We* ~ 0.3 and *Ca* ~ 0.03. Larger bubbles/drops flatten in Newtonian fluids order to reduce the drag. In non-Newtonian fluids, they also flatten in shear thinning fluids, whereas if normal stresses are important, the bubble/drop can take a tear shape with a sharp tip at the end (Figure 4) [1]. In the following, we will not consider the case of large and deformed objects.

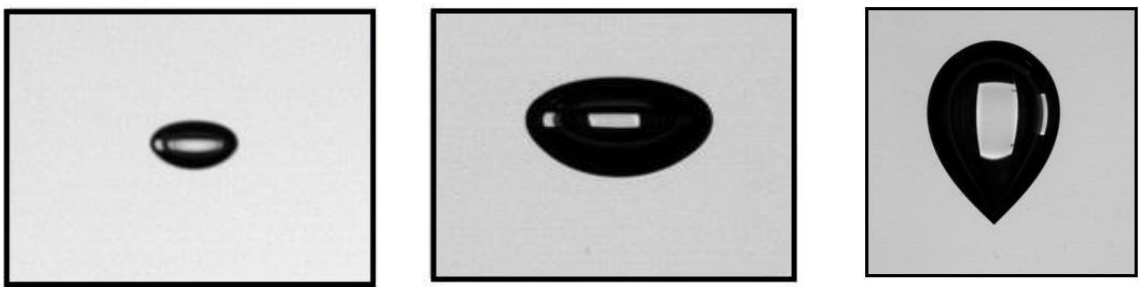

Figure 4. Shape of bubbles rising in a) water, largest diameter $d_m$ = 1.6 mm; b) sugar syrup, $d_m$ = 3.4 mm; c) carboxymethylcellulose (100 ppm in water), $d_m$ =7 mm. Source: Ref [22]

A second question concerns the boundary condition at the bubble/drop surface, which would change the rise velocity. In the absence of surface-active agents, one would expect that the interface is fully mobile, in which case the friction coefficient f is equal to $2\pi\eta d$ at small Reynolds number. In practice, this condition is rarely observed when the dispersion fluid is water, because of contamination by residual surface-active species. Only when water is thoroughly purified, the mobile surface condition applies, a feature that can be used to monitor adsorption of slowly adsorbing surface active species by measuring the time variation of the rising velocity of bubbles [23].

As discussed by Maldarelli and co-workers, the friction coefficient depends on surfactant adsorption and desorption rates [21]. If the surfactant is irreversibly adsorbed, it is swept to the trailing pole of the bubble where it accumulates and lowers the surface tension relative to the front end. The difference in tension creates a Marangoni force which opposes the surface flow and increases the drag coefficient. However, if the surfactant can exchange between bulk and surface, new surfactant can adsorb from the bulk and replenish the front surface, and the surfactant accumulated at the rear can desorb; as a result, the surface tension gradients disappear and the Marangoni force can be suppressed. Maldarelli *et al* considered the case of dilute surfactant solutions and small Reynolds numbers with kinetically or diffusion limited sorption. They showed that the Marangoni force is present if the rate of either kinetic or diffusive transport of surfactant to the bubble surface is slow relative to surface convection and if surface diffusion is also slow. They performed experiments with dilute solutions of surfactants and the results are shown in Figure 5: the surfactant is unable to immobilize the surface of the bubbles only if its concentration is extremely small, well below the the critical micellar concentration (*cmc*).

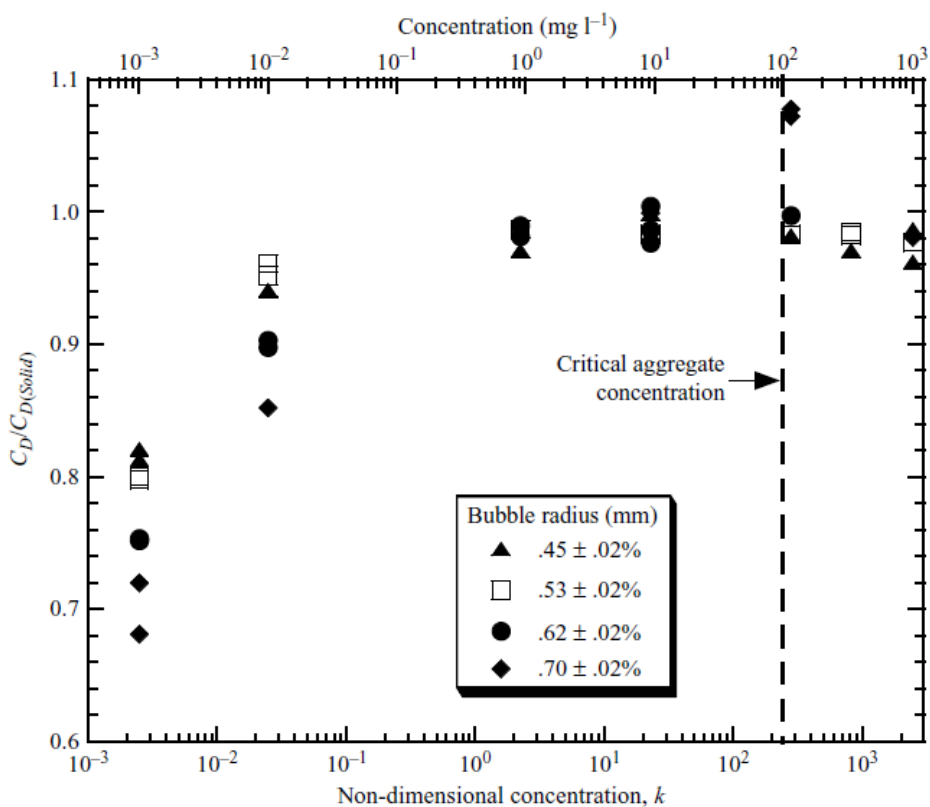

Figure 5. Effect of the concentration of a nonionic surfactant ($C_{12}E_6$) on the drag coefficient (divided by the drag coefficient for a solid sphere $C_D$(Solid) at the same *Re)*. Reprinted with permission from [21]. Copyright 2006 Cambridge University Press.

Maldarelli *et al* also predicted that the Marangoni force is larger for smaller bubbles. Above the *cmc*, the adsorption kinetics is also controlled by the micelle lifetime, which can very long (minutes) with nonionic surfactants and be rate limiting.

Counterintuitively, the effect of surfactants can be less important than that of trace contaminants because their concentration is larger and exchanges between surface and bulk are faster. This phenomenon was observed it during the flow of bubbles in capillaries and called surfactant remobilization by Stebe and Maldarelli. They found that remobilization occurred at surfactant concentrations close to the critical micellar concentration (*cmc*) and pointed out that it occurs when the kinetics of micellar breakup which is the source of surfactant monomers is fast compared to the surface convective flux [24].

A third difference between bubbles/drops and solid particles is the change in bubble/drop size that can occur due either to coalescence or to Ostwald ripening (transfer of gas or of the drop's liquid due to pressure differences between bubbles/drops of unequal size). Ostwald ripening is sometimes called disproportionation for bubbles and coarsening for foams.

Studies of bubbles and drops are therefore difficult to interpret, explaining why the behavior of solid particles is better understood. However, much has be learned from studies of solid particles

### 4.1. Submicronic drops and bubbles

Submicronic drops or bubbles can have sizes smaller than the characteristic size of the medium in which they move. Ostwald ripening is usually very fast for bubbles, because it depends on the gas solubility in the liquid which is much larger than the mutual solubility of many oils and water. In practice, it is quite difficult to study bubbles with radius smaller than a few tens of microns. Submicronic drops are easier to produce, in turbulent devices for instance [25].

Very small bubbles/drops can however be produced by nucleation. In a recent study, Rosowski *et al* observed that these small bubbles can be trapped in polymeric networks when their size reaches the mesh size of the network and when the elastic modulus is larger than the pressure in the drop: the network squeezes the drop [26]. However, as long as the pressure within a droplet exceeds a critical value $P_c$, it can overcome elastic stresses and continue to grow. In this work, it was assumed that $P_c$= 5E/6, E being the Young modulus, a classical result for nonlinear-elastic solids, which is a good approximation for many polymer networks. Note that $E = 2G (1+\nu)$, $\nu$ being the Poisson coefficient ($\nu$ = 0.5 for incompressible media, i.e. for most fluids and elastomers).

The authors prepared samples with a gradient in stiffness and observed that drops move to the soft side and where they continue growing. The phenomenon is analog to Ostwald ripening, but much faster. It could play a role in biology: in the nucleus of living cells, artificial phase-separating domains were found to form preferentially in softer chromatin-poor regions. After drops were triggered to grow in chromatin-rich regions, they migrated toward chromatin-poor regions [27].

Microrheology methods were recently used by Mason *et al* to probe the rheology of emulsions with monodisperse nanoemulsion drops. In these experiments, the drops play both the role of tracer particles and of jammed and deformed objects imparting elasticity to the medium [25]. Classical microrheology methods rely on the analysis of the Brownian motion of isolated particles in the medium. The Brownian motion of drops in concentrated emulsions is stronly affected by the other drops. Mason *et al* showed that provided the diffusion coefficient is corrected by using the structure factor of the dispersion, a very good correlation with the rheological parameters determined with standard rheometers is achieved.

### 4.2 Non-Brownian drops and bubbles. Motion under gravity forces.

When the drops/bubbles are larger, bubble rise due to gravity and drops either rise or sediment depending on the density difference with respect to the continuous phase. The rising motion of drops is frequently called creaming, by analogy with the rise of oily drops in milk. As mentioned earlier, if they are not too big (typically diameter below 1mm), they do not deform during the motion and in general behave as solid spheres. For instance, in bubbly liquids, bubbles form strings as solid spheres [1].

For bubbles of diameter 1 mm ascending in a viscous non-Newtonian fluids with viscosity 10,000 times larger than water, the rising velocity can be estimated using the Stokes expression (Equation 6): $V_{St}$ ~ 50 µm/s and $V_{St}/d$ ~ 0.05 s$^{-1}$. At these low shear rates, shear thinning is frequently evidenced and the viscosity to be used in Equation 6 needs a careful evaluation. Approximate solutions were developed and found in agreement with experiment [28]. It is currently considered that the viscosity controlling the rising velocity is the viscosity at a shear rate $\dot{\gamma}$, of the order of V/d, , V being the actual velocity [29].

The boundary condition at the bubble/drop surface has been extensively investigated in Newtonian fluids, in particular water, to which small amounts of surfactant was added to mimic the role of contaminants. It was found that the surfaces remained immobile. Similar investigations were performed with non-Newtonian fluids and led to the same conclusions [28]. In fluids exhibiting non-zero normal stresses, a sharp increase in velocity is frequently observed above a critical diameter. This is sometimes attributed to a transition toward a regime where the surface become mobile. However, the transition could also be due to a shape change of bubbles/drops. The problem is not fully clarified yet [1].

In foams, the bubbles are closely packed and distorted into polyhedral shapes [10]. In the foam structure, the bubbles are flattened and separated by thin liquid films; these films are connected through liquid channels called Plateau borders (PB), themselves interconnected in nodes. Most of the liquid is contained in the PBs and in the nodes, because the liquid films are very thin (thickness of the order of 100 nm) and contain little water. The bubbles rise due to gravity, and the liquid drains through the PB network. In drainage theories, it is either assumed that the bubbles surface is mobile or immobile. Mobile conditions were found with surfactants solutions in concentrations above *cmc*, and immobile conditions with irreversibly adsorbed surface-active species such as proteins, or mixtures of surfactants with cosurfactants in small concentrations.

It is to be noted that foams are usually made with surfactant solutions in concentration well above *cmc*, to ensure proper bubbles surface coverage. During foam drainage, the surfactant layer density at the surface of the bubbles is not affected, because surfactant molecules rapidly replenish the surface if depleted in surfactant. But the bubble surfaces are sheared. It was observed that by increasing the viscosity of the foaming liquid, the boundary condition could change from immobile to mobile. This can be understood introducing the Bousinessq number Bo:

$$Bo = \frac{\eta_s}{\eta d} \qquad (11)$$

When Bo decreases, the dissipation in the bulk liquid dominates over surface dissipation. The same change in surface boundary conditions was observed when increasing the bubble size [30].

Drainage studies in polymer solutions showed that as for single bubbles, the velocity of drainage is controlled by the viscosity value for a shear rate of the order of $V/d_{PB}$, where $d_{PB}$ is the diameter of the Plateau borders [30]. Studies were also made with solutions of polymers with high molecular weight.In this case, the results could not be rationalized using the shear dependent viscosity, and the elongational deformation rates were too small for normal stresses to play a role. It was argued that the fluid is submitted to expansions and contractions when flowing through the nodes, in which case, elongational effects are known to manifest at smaller elongational rates.

When the continuous phase is predominantly elastic, buoyant bubbles/drops can be trapped and stay immobile. This happens when the buoyancy force par unit area is balanced by the elastic stress.

The buoyancy stress increases with the diameter d, and if it is larger than the yield stress of the continuous phase, the bubbles/drops are not immobilized, a condition that reads approximately:

$$\sigma_{y\,bulk} > \frac{\Delta \rho\, g\, d}{6} \qquad (12)$$

in agreement with results for bubbles in polymer solutions [31]. Equation 12 also applies to foams made either from particle dispersions [32] or from viscoelastic lamellar phases [33]: the bubbles are initially small and immobilized, but they grow in time because of the coarsening process; at a certain point the buoyancy stress becomes comparable to the yield stress and the foams start to drain.

A related but more complex mechanism of foam drainage has been reported for dispersions of colloidal clay, Laponite in aqueous SDS solutions [34]. During foam drainage, it was assumed that the interstitial fluid gels because the yield stress of Laponite dispersions increases upon confinement. Drainage is arrested after a time decreasing with increasing Laponite concentration. Due to coarsening, the bubble size increases, the Plateau borders size increases as well, and drainage starts again at a later time. Note that confinement effects have been observed in other gels: the gelation time decreases appreciably when the fluid is confined in spaces much larger than their mesh size [35]. These confinement effects are still poorly understood, but clearly deserves more investigations.

Foams may also contain oil drops when they are made from emulsions [2]. At small oil volume fractions, foamed emulsions drain like foams made from surfactant solutions only. At larger oil volume fractions, very different features are observed. When $\phi > \phi^* = 0.64$ (random close packing of spheres), emulsion droplets are densely packed, the emulsions become viscoelastic and have a finite shear modulus and yield stress. Emulsion drops are actually confined and crowded between bubbles, which stay anomalously far from each other. The presence of such a dense assembly of droplets trapped and jammed in between bubbles has several effects. The local viscosity increases, slowing down both film thinning and Plateau border shrinking (slower drainage). The yield stress may also increase upon confinement as in the Laponite-SDS foams. Buoyancy stresses in Plateau borders may be less than the emulsion yield stress and drainage can be arrested.

In experiments performed by foamed concentrated emulsions, it was found that whereas the yield stress of the interstitial fluid stabilizes the foam at rest, rapid drainage is induced by shear [36].

Note that, as gravity motion, ripening can be stopped with a gelified continuous phase. In this case, and as seen in section 4.1 for submicronic bubbles, the Young modulus should be larger than the capillary pressure, itself much larger than the buoyancy pressure, especially in emulsions. For dilute dispersions of bubbles/drops, the maximum radius above which the dispersion ripen is such that the capillary pressure is of the order of the elastic modulus [26]. For concentrated dispersions, the capillary pressure is higher and ripening is less easy to arrest. Detailed calculations for foams lead to a relation between elastic modulus and capillary pressure, in good agreement with experiments [37].

### 4.3 Rheology of non-Newtonian fluids containing bubbles or drops

In predominantly viscous matrices, and at small shear rates (when the bubbles/drops are not deformed), the shear viscosity increases with the volume fraction of bubbles and drops, as already seen for solid particles (Equation 3). The normal stresses and the extensional viscosity increase as well [38].

The elastic modulus is also affected by the presence of bubbles/drops and differ from that of the continuous phase. The mechanical strength of solids with elastic modulus much larger than the capillary pressure is mainly affected by voids in the bulk matrix. Liquid inclusions which have no shear modulus reduce the stiffness of a solid composite, while solid inclusions increase this stiffness. Softer elastic matrices can have elastic moduli closer to the capillary pressure, for instance polymer gels, cross-linked networks of actin or gelatin, concentrated emulsions, dense granular suspensions and pastes. In those materials, small liquid drops can either stiffen or soften a polymer matrix [39].

Most studies were focused on systems for which the volume fraction of capillary inclusions is smaller than about 0.64, the packing volume fraction of spherical inclusions. Recent studies were performed by Pitois *et al* with foams made with concentrated emulsions [40]. The elastic modulus is governed by two parameters: the gas volume fraction and the elasto-capillary number :

$$Ca_{el} = \frac{R}{G\gamma} \qquad (13)$$

G being the shear modulus of the matrix. The results together with those for small bubble volume fractions are shown in Figure 6 [40].

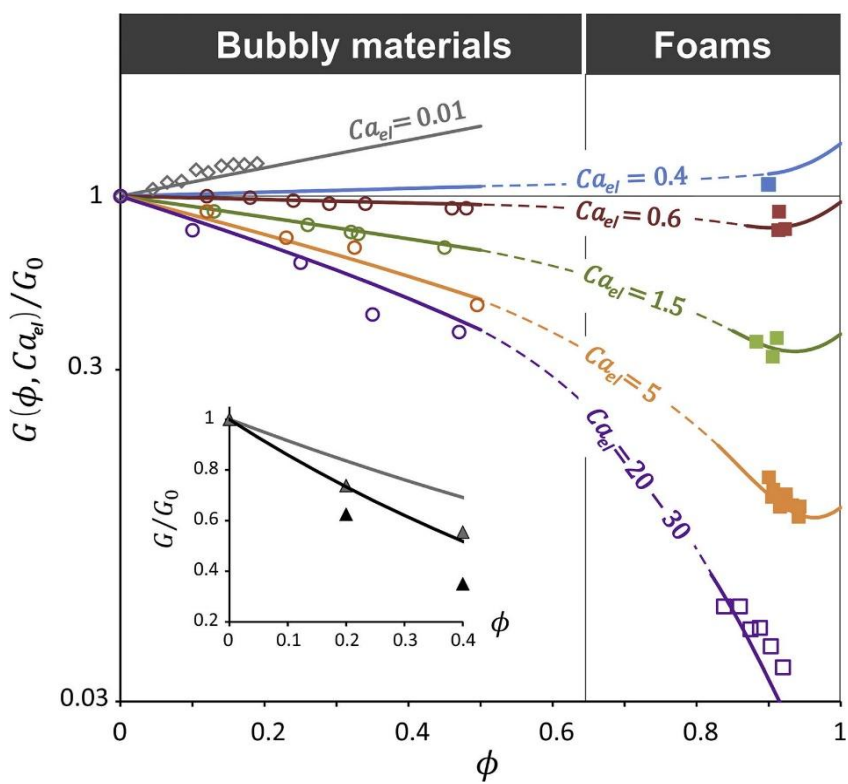

Figure 6. Ratio of shear elastic modulus of bubbly and foamy solids divided by the elastic modulus of the bulk material as a function of gas volume fraction , for several values of the elasto-capillary number . Reprinted with permission from [40]. Copyright 2017 Elsevier

## 5. Conclusions

The motion of solid particles in quiescent Newtonian fluids is now quite well understood, but open questions remain in relation to their motion in non-Newtonian fluids, in particular when their concentration is important or when normal stresses play a role. The case of bubbles and drops is still less well understood, although significant progress has been made during the recent years. The main difficulties lie in the facts that they are deformable, the surface conditions are rarely known and their size may change due to Ostwald ripening or coalescence.

Deformations can be different than in Newtonian fluids, but these aspects have not been discussed here. The question of surface boundary conditions has been extensively investigated in Newtonian fluids, in connection with the role of contaminants. Most of the works are thus focused on solutions containing very small concentration of surfactant and on bubbly liquids. Very little work has been performed in non-Newtonian fluids, even with dilute surfactant solutions. What happens above *cmc* is still unclear and need to be investigated.

Interesting specific effects has been evidenced, such as drop/bubble arrest, including in concentrated emulsions and foams when the yield stress of the liquid matrix is large enough. Arrest of Ostwald ripening was also observed when the elastic modulus of the matrix is larger. A new type of ripening was observed in systems with inhomogeneous elastic properties.

The rheology of non-Newtonian systems containing drops or bubbles is another active field. Stiffening of softening can be observed, depending on the ratio between capillary pressure and elastic modulus of the matrix, and on the volume fraction of the inclusions.

All these studies have important consequences for practical applications and more investigations are needed to improve the current understanding.

## Acknowledgements

I am grateful to Annina Salonen for very stimulating discussions and to Kate Stebe for useful suggestions concerning the hydrodynamic boundary condition at bubbles or drop surfaces.